\documentclass[conference]{IEEEtran}
\IEEEoverridecommandlockouts
\usepackage{cite}
\usepackage{amsmath,amssymb,amsfonts}
\usepackage{algorithmic}
\usepackage{graphicx}
\usepackage{textcomp}
\usepackage{makecell}
\usepackage{algorithm}
\usepackage{multirow}
\usepackage{graphicx}      
\usepackage{subcaption}
\usepackage{pgfplots}
\usepackage{url}
\usepackage{tikz}
\usepackage{hyperref}
\usepackage{comment}
\usepackage{xcolor}
\def\BibTeX{{\rm B\kern-.05em{\sc i\kern-.025em b}\kern-.08em
    T\kern-.1667em\lower.7ex\hbox{E}\kern-.125emX}}
\begin{document}

\title{\textit{SecureFed}: A Two-Phase Framework for Detecting Malicious Clients in Federated Learning}

\author{\IEEEauthorblockN{Likhitha Annapurna Kavuri\IEEEauthorrefmark{1}, Akshay Mhatre\IEEEauthorrefmark{2}, Akarsh K Nair\IEEEauthorrefmark{3}, Deepti Gupta\IEEEauthorrefmark{4}}
\IEEEauthorblockA{\IEEEauthorrefmark{1}\IEEEauthorrefmark{2}\IEEEauthorrefmark{4}Dept. of Computer Information Systems, Texas A\&M University - Central Texas, Texas, USA \\\IEEEauthorrefmark{3}{Department of CSE, IIIT Kottayam, Kottayam, India}\\}
\IEEEauthorrefmark{1}lk040@my.tamuct.edu, 
\IEEEauthorrefmark{2}am271@my.tamuct.edu,
\IEEEauthorrefmark{3}akarshkn@iiitkottayam.ac.in,
\IEEEauthorrefmark{4}d.gupta@tamuct.edu}




\maketitle

\begin{abstract}

Federated Learning (FL) protects data privacy while providing a decentralized method for training models. However, because of the distributed schema, it is susceptible to adversarial clients that could alter results or sabotage model performance. This study presents \textit{SecureFed}, a two-phase FL framework for identifying and reducing the impact of such attackers. Phase 1 involves
collecting model updates from participating clients and applying
a dimensionality reduction approach to identify outlier patterns
frequently associated with malicious behavior. Temporary models
constructed from the client updates are evaluated on synthetic datasets 
to compute validation losses and support anomaly scoring. The idea 
of learning zones is presented in Phase 2, where weights are dynamically 
routed according to their contribution scores and gradient magnitudes. 
High-value gradient zones are given greater weight in aggregation and 
contribute more significantly to the global model, while lower-value 
gradient zones, which may indicate possible adversarial activity, are 
gradually removed from training. Until the model converges and a strong 
defense against poisoning attacks is possible, this training cycle continues 
Based on the experimental findings, \textit{SecureFed} considerably improves model
resilience without compromising model performance.

\end{abstract}


\section{Introduction}

Federated Learning (FL) is a decentralized machine learning paradigm that enables multiple clients to collaboratively train a shared global model while maintaining their raw data local~\cite{bonawitz2019towards}. This design provides significant privacy advantages, especially in domains such as healthcare, finance, and mobile applications, where sensitive data cannot be centrally aggregated due to privacy regulations like GDPR. Despite these benefits, FL is highly susceptible to adversarial attacks due to the lack of centralized control over individual client updates~\cite{kairouz2021advances}. One of the major concerns in FL is the presence of malicious clients that send tainted or modified model updates to the server, aiming to degrade the overall model performance or generate targeted misclassifications~\cite{bagdasaryan2020backdoor}. Since malicious updates are usually designed to statistically mimic benign behavior, such 
attacks can be challenging to identify. Therefore, protecting FL systems against similar threats is crucial for preserving their effectiveness and trustworthiness~\cite{sun2020backdoor}.

Motivated by these challenges, we propose \textit{SecureFed}, a two-stage FL framework designed to isolate and detect rogue clients. Initially, \textit{SecureFed} leverages publicly available datasets to collect model updates from all participating clients. It then applies a dimensionality reduction techniques, Principal Component Analysis (PCA) to analyze weight vector patterns \cite{jolliffe2002pca}. This analysis enables us to identify irregularities in client behavior. Further to assist in extensive anomaly detection, a temporary model is constructed using the gradients and validated on a synthetic datasets. The observed error rates and loss values computed from the reduced-dimensional representations further assist in the anomaly detection. This score is carry forwarded to phase 2, where \textit{SecureFed} introduces the concept of learning zones, the zones group clients based on trustworthiness to guide training decisions, quantified using weight magnitudes and gradient values. Clients exhibiting low-gradient patterns, usually associated with adversarial manipulation, are progressively removed from the training loop. In contrast, clients with high-gradient values are deemed reliable and continue contributing to the global model \cite{zhang2022gradient}. The proposed zoning strategy enables targeted learning and enhances resistance to poisoning attacks.

The contributions of the article are as follows:
\begin{itemize}
\item A federated anomaly detection framework is proposed, leveraging dimensionality reduction to identify suspicious patterns in client weight updates.
  
\item A preliminary scoring model is developed, trained on real-world attack datasets to support early-stage anomaly detection.
   \item A two-stage architecture termed \textit{SecureFed} is presented, combining client scoring and trust-based filtering.  
\item A dynamic learning zone mechanism is introduced, adjusting the model training process based on client trustworthiness.
\item Extensive experiments are conducted 
to demonstrate the effectiveness of \textit{SecureFed} in mitigating adversarial impacts.

\end{itemize}

The remainder of the article is structured as follows: Section~\ref{Related} reviews the
background and relevant FL security. Section~\ref{problem} outlines the problem formulation followed by the introduction of \textit{SecureFed} framework in Section~\ref{proposed}.  Section~\ref{results} presents the experimental setup, analysis, and results. Finally, Section~\ref{diss} presents the discussions and Section~\ref{conclusion} concludes the paper and presents future research directions.
\section{Background and Related Work}\label{Related}
This section reviews the existing literature on security threats in FL systems and highlights defense mechanisms developed to detect and mitigate malicious client behavior.
\subsection{Security Issues in Federated Learning}

FL was introduced as a privacy-preserving training paradigm~\cite{rathee2023elsa}. However, 
FL systems are susceptible to adversarial threats, including poisoning, leakage and backdoor attacks. In poisoning attacks, malicious clients injects adversarial data into the system, degrading model accuracy and sabotaging system performance. These attacks ultimately aim to shift the model's focus toward adversarial tasks and remain undetected to high extent. Backdoor attacks employ a more subtle approach, targeting specific inputs to trigger the model into executing adversarial behavior. These backdoors are stealthily embedded into the system in such a way that the model’s performance on the primary task remains unaffected, allowing the adversarial activity to go undetected~\cite{sun2020backdoor}.
Additionally, the exchange of gradients or model weights during training may leak sensitive information (unintentionally or even intentionally), exposing client-side data. These leaks can be exploited by adversaries to launch powerful data leakage and reconstruction attacks. In such attacks, adversaries are able to reconstruct input data from the leaked information, effectively gaining access without directly participating in the system. Detecting and mitigating these threats remains a critical challenge in federated systems.
Since malicious updates often mimic benign ones, traditional outlier-based defenses are often ineffective. This emphasizes the need for robust, adaptive defense mechanisms that can identify and isolate malicious behavior without disrupting the overall training process.

\subsection{Methods to Detecting Malicious Clients in FL}

Zhang et al. \cite{zhang2022challenges} presented one of the initial comprehensive reviews that discussed the challenges in FL, highlighting vulnerabilities such as model poisoning, data leakage, and high communication costs was presented by Zhang et al.\cite{zhang2022challenges}. The article emphasizes the complexity of ensuring both privacy and robustness in decentralized settings. It evaluates existing defense techniques like differential privacy, homomorphic encryption, and secret sharing, while identifying gaps in incentive mechanisms and personalization. This work presents the broader problem space where several FL security frameworks similar to \textit{SecureFed} operate.

Li et al. \cite{li2020learning} presented an article focusing on the detection of malicious clients, proposing a VAE-based framework for learning low-dimensional embeddings of model updates. Malicious updates generate higher reconstruction errors, enabling their detection without labeled data. The method works in both unsupervised and semi-supervised modes with dynamic thresholds. 
The framework emphasizes adaptive, data-driven detection of harmful contributions for both targeted and untargeted attacks.
Similarly, FedDMC~\cite{mu2024feddmc} is another framework focusing on identifying malicious clients in poisoning attack settings. The framework combines Binary Tree-Based Clustering with Noise, PCA, and a Self-Assemble Detection Correcting Module, forming a poisoning-resistant FL system. The system detects malicious clients without clean validation data by balancing high detection accuracy and low computational overhead. 
FedDMC's multi-tiered detection strategy has similarities to \textit{SecureFed}'s use of dimensionality reduction and trust-based filtering.

Gupta et al. \cite{gupta2021hierarchical} have proposed an FL-based anomaly detection framework for healthcare systems. The framework combines Digital Twins and Edge Cloudlet Computing to avoid sensitive data transfer. In addition, several security models for protecting IoT devices are discussed in~\cite{praharaj2023hierarchical,karim2025securing}.

Even though the primary focus is not on adversarial detection, the work highlights FL's potential in secure and privacy-preserving applications.

\subsection{Comparison with existing works}

To improve robustness against adversarial updates, several defenses rely on similarity or distance-based heuristics. Conventional mechanisms like Krum selects updates closest to the majority~\cite{blanchard2017machine}, while statistical aggregators like the median and trimmed mean offer resilience to outliers but overlook deeper structural patterns in client behavior~\cite{yin2018byzantine}. FoolsGold~\cite{fung2020limitations} targets Sybil attacks by analyzing gradient similarity, but may struggle with more nuanced poisoning attempts. FLAME~\cite{ma2022flame} is another popular mitigation strategy that uses adaptive weighting and gradient clipping.  In contrast, \textit{SecureFed} leverages dimensionality reduction to detect anomalies based on the underlying distribution of client updates, combining it with a trust-aware approach that dynamically filters clients based on anomaly and gradient-based scoring.

\textit{SecureFed} differs from FLTrust~\cite{cao2020fltrust}, which uses a static reference model without revalidating client behavior, by incorporating synthetic validation prior to aggregation. Unlike prior works that apply dimensionality reduction solely for post-hoc analysis~\cite{zhao2021dpPCA}, \textit{SecureFed }integrates it directly into the FL pipeline to enable iterative anomaly detection.

\section{Problem Identification}\label{problem}

Even though FL enhances privacy by keeping data local, its decentralized architecture introduces significant security vulnerabilities. Particularly, malicious clients may inject poisoned updates that degrade model performance or embed backdoors into the global model.

\subsection{Vulnerabilities in Federated Learning}

FL is vulnerable to adversarial clients whose poisoned updates can degrade global model performance. Traditional aggregation methods such as Federated Averaging (FedAvg) are vulnerable to such attacks as they assume all clients to be benign. To address this, several robust aggregation strategies have been proposed. One such framework, FLTrust~\cite{cao2020fltrust} introduces a server-side trust bootstrapping mechanism, where the server maintains a small, trusted dataset. Client updates are scored based on their similarity to this baseline, helping to reduce the impact of malicious clients. MAB-RFL~\cite{wan2022shielding} employs a multi-armed bandit strategy to adaptively select clients for aggregation. By formulating client selection as a bandit problem, the framework balances exploration and exploitation to identify and prioritize reliable clients over time.


However, these defenses face limitations, particularly in high-dimensional settings, where benign anomalies and adversarial manipulations are difficult to distinguish.
\subsection{Dimensionality Reduction for Malicious Client Detection}

High-dimensional model updates in FL can make it challenging to distinguish between benign and malicious behavior. To address this, dimensionality reduction techniques such as autoencoders and PCA have been explored:

\begin{itemize}
    
\item \textbf{Autoencoder-Based Anomaly Detection:} Autoencoders can learn compressed representations of benign client updates by capturing their underlying structure. Deviations from this structure (reflected as high reconstruction errors) usually indicate anomalous or malicious activity.

\item \textbf{Gradient and Reconstruction Analysis:}
Existing literature has proven that combining gradient information with autoencoder-based reconstruction improves anomaly detection accuracy in FL. This hybrid approach leverages both gradient deviations and reconstruction errors to enhance robustness against poisoning attacks~\cite{alsulaimawi2024federated}.

\end{itemize}

However, integrating these techniques into federated settings remains challenging, particularly in ensuring an optimal trade-off between representation reduction, client privacy, and potential model performance degradation.


\subsection{Secure Aggregation Protocols}
Secure aggregation algorithms aim to preserve client anonymity by aggregating model updates without revealing individual contributions. However, the privacy guarantee can unintentionally conceal malicious activity.

\begin{itemize}
    
\item \textbf{ELSA} introduces a secure aggregation scheme that distributes trust management between two non-colluding servers. This design ensures the privacy of client updates while enabling detection of malicious behavior~\cite{rathee2023elsa}.
   
\item  \textbf{SeaFlame} enhances communication efficiency using share conversion and sharing techniques. It reduces communication overhead while maintaining robustness against malicious clients \cite{tang2025seaflame}.
\end{itemize}

These protocols highlight the inherent trade-off between protecting client privacy and retaining the ability to detect and mitigate adversarial activities.

\subsection{Feature Engineering and Client Behavior Analysis}

Feature engineering plays a critical role in enhancing the detection of malicious clients.

\begin{itemize}
    \item \textbf{Feature Selection Techniques:} Methods such as recursive feature elimination, chi-square tests, and mutual information help identify relevant features that differentiate between benign and adversarial behaviors. Effective feature selection reduces dimensionality and improves model interpretability.

    \item \textbf{Modeling Client Behavior:}
Understanding and modeling client behavior can assist in identifying anomalies. By analyzing patterns in client updates over time, certain deviations can be observed, indicative of malicious intent.
\end{itemize}
\subsection{Research Gap}
Even with the advancements in secure aggregation, anomaly detection, and robust aggregation strategies, several key challenges can be identified:
\begin{itemize}
    \item \textbf{High-Dimensional Data:}
Existing defenses often fail in high-dimensional settings, where distinguishing adversarial behavior from benign outliers becomes unreliable.
\item \textbf{Integration of Techniques:}
There is a lack of unified frameworks that effectively combines dimensionality reduction, anomaly detection, and secure aggregation to address malicious client identification.
\item \textbf{Adaptive Learning Zones:} Current systems are primarily static, failing to dynamically adjust to evolving client behavior, limiting their long-term robustness.
\end{itemize}

\subsection{Proposed Approach}
Motivated by these research gaps, the study proposes \textit{SecureFed}, a two-stage framework that combines anomaly detection using dimensionality reduction with adaptive client filtering based on trust scores. The proposed framework aims to strengthen FL against adversarial attacks by improving adaptive security via learning zone based client screening. 
    

    


\section{Proposed Framework}\label{proposed}
The \textit{SecureFed} framework is designed to enhance the robustness of FL systems via a multi stage schema, presented in Figure~\ref{block}. 
It operates through two core phases: an anomaly detection module and an adaptive aggregation module.


\subsection{First layer: Client Side}

To simulate the FL setup and extract client-side model updates for subsequent analysis, the system begin by using publicly available attack datasets. Data is then partitioned in IID settings, followed by standard preprocessing approhces such as reshaping and normalization. After preprocessing, FL training is initiated by transferring the global model to clients, who then begin their initial round of training. After training, clients send their updated model weights to the central server for aggregation and anomaly analysis.

.
\subsection{Second Layer: Server Side}
Server-side processing in \textit{SecureFed} operates in two phases: anomaly detection and adaptive zone-based aggregation.

\subsubsection{\textbf{Phase 1  (Anomaly Detection)}}

After receiving weight updates $\{\mathcal{W}_c^r\}$ from all clients in round $r$, the server applies PCA to reduce each client’s update into a lower-dimensional space:
\[
\tilde{\mathcal{W}}_c^r = \text{PCA}(\mathcal{W}_c^r)
\]

An anomaly score $A_c$ is computed for each client based on deviations in the reduced representation. To calibrate the detection process, a threshold $\tau^*$ is estimated using a synthetic validation dataset $D_s$. For the scope of this work, a standard dataset with similar feature vectors to the training dataset was used~\cite{Dataset} in the place of synthetic data:
\[
\tau^* = \text{Validate}(D_s, \{A_c\})
\]

This threshold is further used to normalize and scale trust computation in Phase~\ref{phase2}.

\subsubsection{\textbf{Phase 2 ( Adaptive Learning Zones)}}\label{phase2}

Each client’s update $\mathcal{W}_c^r$ is temporarily applied to the current global model to form a personalized model $\mathcal{M}_c^r$:
\[
\mathcal{M}_c^r = \mathcal{M}_g^r + \mathcal{W}_c^r
\]

This temporary model is evaluated on the synthetic dataset $D_s$ to calculate a validation loss $L_c$, which helps assess the behavioral consistency of each client’s contribution. Simultaneously, the gradient magnitude is computed as:
\[
G_c = \|\nabla \mathcal{W}_c^r\|
\]

Next, a trust score $T_c$ is calculated by combining the normalized anomaly score, validation loss, and gradient magnitude:
\[
T_c = \alpha \cdot \left(1 - \frac{A_c}{\tau^*} \right) + \beta \cdot \left(1 - \frac{L_c}{\max(L)} \right) + \gamma \cdot \frac{G_c}{\max(G)}
\]

Based on this score, clients are dynamically assigned to one of three learning zones determined by a preset threshhold value: Zone 1 (High Trust) , Zone 2 (Uncertain) and Zone 3 (Low Trust). Each zone is assigned a weighting factor $\alpha_z(c)$ which determines its influence during aggregation. The global model is updated through zone-weighted aggregation as follows:
\[
\mathcal{W}_{r+1} = \frac{\sum_{c \in \mathcal{C}} \alpha_z(c) \cdot n_c \cdot \mathcal{W}_c^r}{\sum_{c \in \mathcal{C}} \alpha_z(c) \cdot n_c}
\]
\[
\mathcal{M}_g^{r+1} = \mathcal{M}_g^r + \mathcal{W}_{r+1}
\]

\begin{figure}[ht]
  \centering
  \includegraphics[height=8cm,width= 7.5cm]{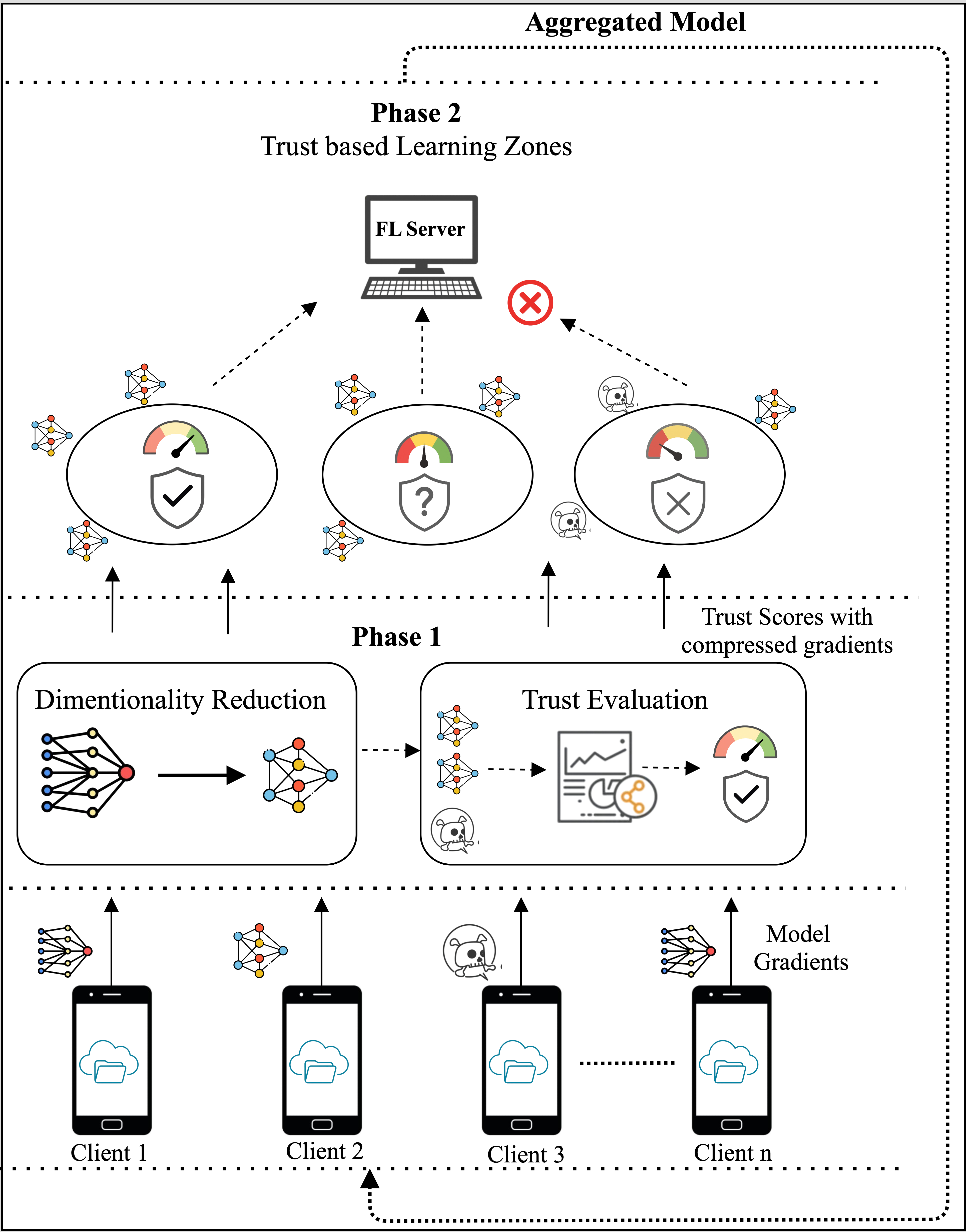}
  \caption{Secure\_Fed : Data flow diagram}
  \label{block}
\end{figure}

The updated model $\mathcal{M}_g^{r+1}$ is then sent to all clients for the next training round. This process continues until model convergence. The overall framework of the proposed system is illustrated in Fig.~\ref{block}. This mechanism ensures that only trustworthy clients contribute to the global model. The detailed working of the framework is presented in Algorithm~\ref{securefed_weighted}.

\begin{algorithm}[ht]
\caption{SecureFed: Two-Phase Framework with Trust-Zone Weighted Aggregation}
\begin{algorithmic}[1]
\REQUIRE Initial global model $\mathcal{M}_g^0$, client set $\mathcal{C}$, public dataset $D_p$, synthetic dataset $D_s$, number of rounds $R$
\ENSURE Final global model $\mathcal{M}_g^R$

\FOR{each round $r = 1$ to $R$}
    \STATE Server broadcasts $\mathcal{M}_g^r$ to all clients $\mathcal{C}$
    \FOR{each client $c \in \mathcal{C}$ in parallel}
        \STATE Client $c$ trains and returns $\mathcal{W}_c^r$
    \ENDFOR

    \STATE \textbf{Phase 1: Anomaly Detection}
    \STATE Collect all client updates: $\{\mathcal{W}_c^r\}$
    \STATE Apply dimensionality reduction: $\tilde{\mathcal{W}}_c^r = DR(\mathcal{W}_c^r)$
    \STATE Compute anomaly scores: $A_c = f_{\text{anomaly}}(\tilde{\mathcal{W}}_c^r)$
    \STATE Estimate detection threshold via synthetic evaluation: $\tau^* = \text{Validate}(D_s, \{A_c\})$

    \STATE \textbf{Phase 2: Adaptive Learning Zones}
    \FOR{each client $c$}
        \STATE Temporarily update model: $\mathcal{M}_c^r = \mathcal{M}_g^r + \mathcal{W}_c^r$
        \STATE Evaluate $\mathcal{M}_c^r$ on synthetic data $D_s$ to compute validation loss: $L_c$
        \STATE Compute gradient magnitude: $G_c = \|\nabla \mathcal{W}_c^r\|$
        \STATE Compute trust score: $T_c = \alpha \cdot \left(1 - \frac{A_c}{\tau^*} \right) + \beta \cdot \left(1 - \frac{L_c}{\max(L)} \right) + \gamma \cdot \frac{G_c}{\max(G)}$

        \IF{$T_c \geq \tau_{\text{high}}$}
            \STATE Assign $c$ to Zone 1 (High Trust), set $\alpha_z(c) = \alpha_1$
        \ELSIF{$\tau_{\text{low}} \leq T_c < \tau_{\text{high}}$}
            \STATE Assign $c$ to Zone 2 (Uncertain), set $\alpha_z(c) = \alpha_2$
        \ELSE
            \STATE Assign $c$ to Zone 3 (Low Trust), set $\alpha_z(c) = \alpha_3$
        \ENDIF
    \ENDFOR

    \STATE \textbf{Weighted Aggregation from All Zones:}
    \STATE $\mathcal{W}_{r+1} \leftarrow \frac{\sum\limits_{c \in \mathcal{C}} \alpha_z(c) \cdot n_c \cdot \mathcal{W}_c^r}{\sum\limits_{c \in \mathcal{C}} \alpha_z(c) \cdot n_c}$
    \STATE Update global model: $\mathcal{M}_g^{r+1} \leftarrow \mathcal{M}_g^r + \mathcal{W}_{r+1}$

\ENDFOR

\RETURN Final global model $\mathcal{M}_g^R$
\end{algorithmic}
\label{securefed_weighted}
\end{algorithm}



\section{Results}\label{results}

\subsection{Experimental Setup}
To evaluate the performance of \textit{SecureFed}, we simulate a FL environment using the standard  MNIST dataset, distributed under IID schema.
The system is trained for three global rounds with 20 clients, with varying malicious client ratios (30-48\%) injecting the model with poisoning updates. The poisoning attack primarily focuses on single-class label flipping mechanism~\cite{2020arXiv200708432T}. Each client trains their model on local data, and the aggregation is initially performed using the vanilla FL schema and then using the proposed \textit{SecureFed} schema. The \textit{SecureFed} schema uses PCA for dimensionality reduction (retaining top-5 components) combined with  K-Means based anomaly clustering, followed by a synthetic dataset based validation for trust-based scoring.
\subsection{Adversarial Robustness Evaluation}
The primary focus of the \textit{SecureFed} framework is attaining adversarial robustness. Thus, the system's ability to isolate malicious clients is measured using precision, recall, and F1-score of the aggregated model. Additionally, the aggregated model's accuracy is also compared against baselines trained under both benign and poisoned environments. As shown in Table~\ref{tab1}, \textit{SecureFed} achieves the highest scores compared to vanilla FL, due to the usage of low-dimensional behavior patterns and synthetic data-based model evaluation.

\begin{table}[h!]
\centering
\caption{Model performance metrics under varying malicious client ratios.}
\label{tab1}
\begin{tabular}{|c|c|c|c|c|c|}
\hline
\textbf{FL Method} & \textbf{Client Status} & \textbf{A (\%)} & \textbf{P} & \textbf{R} & \textbf{F1} \\
\hline
\multirow{3}{*}{\makecell{Vanilla FL\\(FedAvg)}} 
& Benign & 95.49\%& 0.95& 0.95& 0.95\\
& 30\% Malicious  & 92.71\%& 0.91& 0.93& 0.93\\
& 48\% Malicious  & 84.42\%& 0.90& 0.86& 0.84\\
\hline
\multirow{3}{*}{\makecell{SecureFed}} 
& Benign & 95.49\%& 0.95& 0.95& 0.95\\
& 30\% Malicious  & 93.11\%& 0.93& 0.94& 0.93\\
& 48\% Malicious  & 92.50\%& 0.91& 0.92& 0.92\\
\hline
\multicolumn{6}{|l|}{
\small A $\rightarrow$ Accuracy, P $\rightarrow$ Precision, R $\rightarrow$ Recall, F1 $\rightarrow$ F1 Score
} \\
\hline
\end{tabular}
\end{table}

\subsection{Trust Score Dynamics and Zone Distribution}
\begin{figure*}
\centering
 \includegraphics[width=1\textwidth, height=.32\textheight]{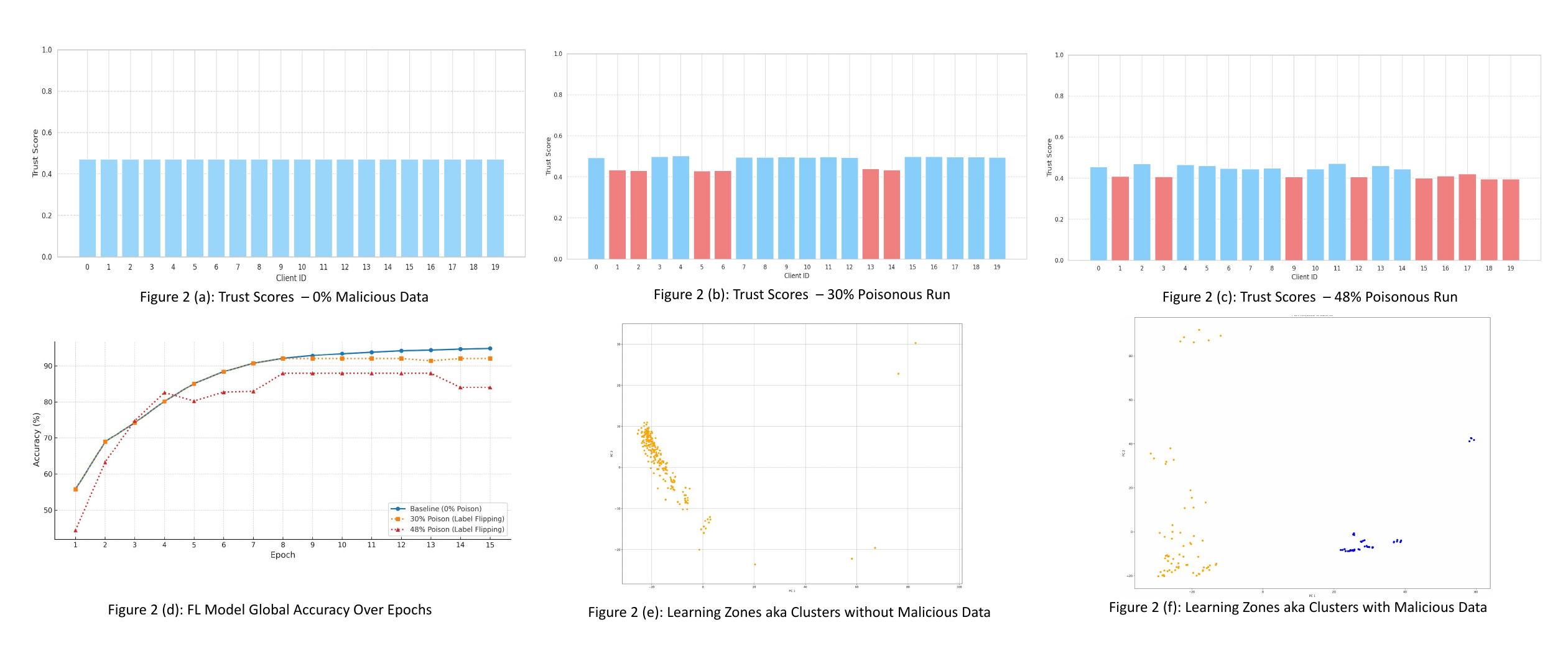}
\centering
\caption{Results}
\label{fig2}
\end{figure*}

Figure~\ref{fig2} illustrates the functioning of trust score based zones and how they evolve over rounds. Benign clients gradually converge to high-trust zones (Zone 1), while malicious ones remain in or migrate to Zone 3. Based on the zones, a weighted averaging is employed, ensuring primary contributions from clients in zone 1. The observations validates the effectiveness of adaptive learning zones in filtering malicious behavior without compromising useful contributions.


\subsection{Ablation Study}
Further, to analyze the contribution of each \textit{SecureFed} component, an ablation study was performed and the observsation are as noted in Table~\ref{ablation}.

\begin{table}[h]
\centering
\caption{Component-wise Ablation Study}
\begin{tabular}{|l|c|c|}
\hline
\textbf{Configuration} & \textbf{Accuracy} & \textbf{Detection Rate} \\
\hline
SecureFed (Full) & 92.50\%& 75\%\\
w/o PCA & 91.61\%& 66.75\%\\
w/o Synthetic Validation & 88.54\%& 38.25\%\\
w/o Trust Score (Binary Filter) & 89.27\%& 45.05\%\\
\hline
\end{tabular}
\label{ablation}
\end{table}

Observations showcase that PCA-based detection, synthetic validation, and trust-driven aggregation together ensure the robustness and reliability offered by the \textit{SecureFed} framework.

\section{Discussion} \label{diss}

The evaluation results demonstrate the effectiveness of the proposed \textit{SecureFed} framework for mitigating the influence of malicious clients while preserving overall model performance. As shown in Table~\ref{tab1}, \textit{SecureFed} consistently outperforms the baseline Vanilla FL (FedAvg) under increasing proportions of adversarial clients. specifically, with 48\% of malicious clients, \textit{SecureFed} achieves an  accuracy of 92.50\%, compared to only 84.42\% with Vanilla FL, an  improvement of 8.08\%. Similarly, the F1 score improves from 0.84 to 0.92, showcasing an increase of 9.5\%, indicating better balance between precision and recall in adversarial scenarios. This highlights the resilience of \textit{SecureFed} in malicious settings.

In scenarios with 30\% malicious clients, \textit{SecureFed} maintains its robustness, improving accuracy from 92.71\% (FedAvg) to 93.11\%, and increasing the F1 score from 0.93 to 0.93 (and a slight increase in precision and recall), indicating that the framework does not compromise performance even when with lesser adversarial clients. The ablation results in Table~\ref{ablation} provide a comprehensive overview of the contribution of each framework component. When PCA-based is removed, the detection rate drops from 75\% to 66.75\%, and accuracy declines by nearly 0.9\%, indicating that PCA plays a moderate but critical role in outlier isolation. A more significant performance degradation is observed when synthetic validation is excluded as accuracy drops to 88.54\% and detection rate reduces by half to 38.25\%. These observations confirms the importance of cross-verification with synthetic data for extensive anomaly detection.

Finally, removing the trust score mechanism and relying solely on binary filters result in a reduced accuracy (89.27\%) and incomplete detection capability, highlighting the value of trust-based aggregation over rigid thresholding. 

Overall, these observations validate the functionality of \textit{SecureFed}. Each module contributes significantly to both performance and defense. The dynamic learning zone strategy enables graceful degradation in adversarial scenarios without abrupt exclusions.


\section{Conclusion and Future work}\label{conclusion}
In this research, we introduced \textit{SecureFed}, a new two-phase architecture that detects and isolates malicious clients to improve the robustness of FL systems.  \textit{SecureFed} incorporates dimensionality reduction techniques and a dynamic learning zone strategy to systematically analyze client inputs and protect the model training process from adversarial participants. 
In Phase 1, we used dimensionality reduction to collect and analyze client weight vectors from publicly available datasets, allowing the early detection of client behavior anomalies.  Further, the idea of learning zones is presented in Phase 2, which enable the framework to filter out possibly dangerous clients based on gradient importance and route reliable updates for further training. Validations on standard attack datasets show that \textit{SecureFed} enhances the federated model's overall performance and security while successfully reducing the effects of poisoned updates.  The method preserves model integrity by dynamically adjusting to adversarial behavior without necessitating major changes to the fundamental FL methodology.

However, a key limit observed in the current framework is its high dependency gradient divergence for anomaly classification. Since the system has only been evaluated under IID data distribution schemes, the underlying assumption holds true. However, in highly divergent settings, the assumption of divergent clients being malicious becomes  less reliable. Thus, in the future work, we propose developing a robust aggregation mechanism that combines gradient divergence with additional behavioral features to support non-IID scenarios. Furthermore, the trust-based filtering mechanism used in \textit{SecureFed} is limited to iteration-specific observations. Thus, in the future works, we plan to formulate an enhanced credibility metrics that integrate trust scores with historical performance trends for more comprehensive and adaptive client evaluation.
\section{Acknowledgment}
This work is partially supported by the US National Science Foundation grant 2431531.

\bibliographystyle{IEEEtran}
\bibliography{References}

\end{document}